# Multipole vector solitons in nonlocal nonlinear media

Yaroslav V. Kartashov, Victor A. Vysloukh*, D. Mihalache** and Lluis Torner

*ICFO-Institut de Ciencies Fotoniques, Mediterranean Technology Park, and Universitat Politecnica de Catalunya, 08860 Castelldefels (Barcelona), Spain*

We show that multipole solitons can be made stable via vectorial coupling in bulk nonlocal nonlinear media. Such vector solitons are composed of mutually incoherent nodeless and multipole components jointly inducing a nonlinear refractive index profile. We found that stabilization of the otherwise highly unstable multipoles occurs below a maximum energy flow. Such threshold is determined by the nonlocality degree.

Nonlocality of the nonlinearity is a property exhibited by many nonlinear optical materials. For example, nonlocality may be important in nematic liquid crystals [1,2], thermal self-action [3], plasmas [4], or photorefractive materials [5]. Nonlocality suppresses modulational instability [6,7]; it stabilizes vortex [8] and two-dimensional fundamental solitons [9]. Since interactions of solitons in nonlocal media are determined by spatial separation, out-of-phase beams could form bound states, a feature predicted in [10] and observed in [11]. The maximum number of solitons that can be packed into stable bound state depends on the nature of nonlocal response [12]. Bound states of dark solitons were addressed in [13,14].

Two-dimensional bright solitons also form bound states in nonlocal media [5,15]. Recently, stable dipole solitons in a medium with Gaussian response function have been predicted [16]. However, many actual materials exhibit nonlocal responses with profiles that depart drastically from Gaussian one. In this Letter we address multipole solitons in a model with nonlocal response of Helmholtz type encountered, in particular, in nematic liquid crystals and plasmas, and show that the shape of nonlocal response is *crucial* for stability of two-dimensional bound states. Thus, *all* scalar bound states are found to be *unstable*, but they can be stabilized *via vectorial coupling* with nodeless soliton. Notice that multipole vector solitons were also studied in the local saturable or photorefractive media [17-25].

We consider propagation of two mutually incoherent laser beams along the $\xi$ axis in media with a nonlocal focusing nonlinearity described by the system of equations for light field amplitudes $q_{1,2}$ and nonlinear contribution to refractive index $n$:

$$i\frac{\partial q_{1,2}}{\partial \xi} = -\frac{1}{2}\left(\frac{\partial^2}{\partial \eta^2} + \frac{\partial^2}{\partial \zeta^2}\right)q_{1,2} - q_{1,2}n,$$
$$n - d\left(\frac{\partial^2}{\partial \eta^2} + \frac{\partial^2}{\partial \zeta^2}\right)n = |q_1|^2 + |q_2|^2, \qquad (1)$$

where $\eta, \zeta$ and $\xi$ stand for the transverse and the longitudinal coordinates scaled to the beam width and diffraction length, respectively. The parameter $d > 0$ stands for the nonlocality degree of the nonlinear response. When $d \to 0$ Eqs. (1) reduce to Manakov vector nonlinear Schrödinger equations; the case $d \to \infty$ corresponds to strongly nonlocal regime. Under proper conditions Eqs. (1) describe nonlocal nonlinearities of partially ionized plasmas resulting from many-body interactions [4] and orientational nonlinearity of nematic liquid crystals (see [1,2] for details of derivation). Eqs. (1) conserve the energy flow $U = U_1 + U_2 = \int\int_{-\infty}^{+\infty}(|q_1|^2 + |q_2|^2)d\eta d\zeta$. The nonlinear contribution to the refractive index is given by $n(\eta,\zeta) = \int\int_{-\infty}^{+\infty} G(\eta - \lambda, \zeta - \tau)(|q_1(\lambda,\tau)|^2 + |q_2(\lambda,\tau)|^2)d\lambda d\tau$, where the response function $G(\eta,\zeta) = (2\pi d)^{-1}\,\mathrm{K}_0[d^{-1/2}(\eta^2 + \zeta^2)^{1/2}]$ is expressed in terms of zero-order MacDonald function. Notice, that in contrast to Gaussian response function of Ref. [16], $G(\eta,\lambda)$ has a logarithmic singularity at $\eta^2 + \zeta^2 \to 0$ and decays slowly. From now on, we will refer to it as Helmholtz response function.

We searched for soliton solutions of Eqs. (1) numerically in the form $q_{1,2}(\eta,\zeta,\xi) = w_{1,2}(\eta,\zeta)\exp(ib_{1,2}\xi)$, where $w_{1,2}$ are real functions, and $b_{1,2}$ are propagation constants. The standard relaxation method was employed that allows to obtain soliton profiles with high accuracy (the difference between calculated profiles for subsequent iterations can be made less than $10^{-16}$). In the scalar case ($w_1 \equiv 0$, $w_2 \neq 0$) we found a variety of solutions composed of several (or single) bright spots with opposite phases that are arranged in rings. Such multipole solitons exist in nonlocal media because the refractive index change in the overlap region between neighboring spots is determined by the total intensity distribution in the entire transverse plane and under proper conditions equilibrium configurations of out-of-phase beams are possible. However, we found that all two-dimensional scalar multipole solitons corresponding to the Helmholtz response are unstable and only nodeless solitons can be stable, in contrast to findings reported in Ref. [16] for Gaussian nonlocal response.

Thus, a first central result of this Letter is that *the physical nature of the nonlocal response is crucial for stability of higher-order solutions in bulk geometries.* Second, we found that *vectorial coupling may lead to stabilization of the vector multipole solitons* even for realistic Helmholtz nonlocal response. Such multipole vector solitons with $w_1 w_2 \neq 0$ were found at $b_2 \leq b_1$ (further without loss of generality we set $b_1 = 3$ and vary $b_2$ and $d$). Profiles of simplest vector solitons are shown in Fig. 1. We do not depict here field $w_1$ of first nodeless component, but instead show the total refractive index profile $n$ and modulus of multipole component $w_2$. With growth of nonlocality degree $d$ the width of refractive index distribution remarkably increases, so that for $d \gg 1$ it greatly exceeds the width of intensity distribution. The total energy flow is a monotonically growing function of propagation constant $b_2$ (Fig. 2(a)). At fixed $b_1$ and $d$ there exist lower $b_2^{\text{low}}$ and upper $b_2^{\text{upp}}$ cutoffs, so that vector solitons can be found only for $b_2^{\text{low}} \leq b_2 \leq b_2^{\text{upp}}$. When $d \to 0$ one has $b_2^{\text{upp}} \to b_1$. With growth of nonlocality degree the width of existence domain for vector solitons shrinks (Figs 2(b) and 2(e)) At the upper cutoff $w_1$ vanishes and vector solitons transform into scalar multipoles; at the lower cutoff $w_2$ vanishes and one gets scalar nodeless soliton. This is illustrated in Fig. 2(d) showing energy sharing $S_{1,2} = U_{1,2}/U$ between dipole soliton components versus $b_2$.

In quasi-local medium ($d \ll 1$) and $b_2 \to b_2^{\text{upp}}$ multipole vector solitons transform into several well-separated monopole vector solitons (the number of solitons is equal to the number of bright spots in $w_2$ field) with weak in-phase $w_1$ and strong out-of-phase $w_2$ components that are both nodeless. In this case the refractive index distribution features several well-separated peaks. When $b_2 \to b_2^{\text{low}}$ strong $w_1$ component remains well localized, while weak $w_2$ component remarkably broadens. In strongly nonlocal medium ($d \gg 1$) both $w_1$ and $w_2$ components are well localized in cutoffs. The refractive index distribution is bell-shaped at $b_2 \to b_2^{\text{low}}$, but at $b_2 \to b_2^{\text{upp}}$ small peaks whose positions coincide with position of intensity maxima in $w_2$ component are clearly observable in otherwise smooth and wide refractive index profile. Nodeless component also features smooth bell-like shape at $b_2 \to b_2^{\text{low}}$ with maximum at $\eta, \zeta = 0$, while as $b_2 \to b_2^{\text{upp}}$ secondary peaks gradually appear in the positions corresponding to intensity maxima in $w_2$ component. Notice, that the presence of peaks in refractive index, even in strongly nonlocal regime, is a specific feature of Helmholtz response in comparison with Gaussian response, where refractive index profile created by multipole beam can be bell-shaped in strongly nonlocal case.

The transformation of vector soliton into stable scalar nodeless beam at $b_2 \to b_2^{\text{low}}$ and unstable multipole beam at $b_2 \to b_2^{\text{upp}}$ suggests the existence of stability domain for composite vector states near $b_2^{\text{low}}$. We found that vector solitons become stable when the energy flow carried by multipole component decreases below certain threshold, i.e., in the region $b_2^{\text{low}} \leq b_2 \leq b_2^{\text{cr}}$. The critical value $b_2^{\text{cr}}$ increases monotonically with $d$ (Fig. 2(c) and 2(f)). Notice that at $d \to 0$, the topological structure of the instability domain becomes complex, so that in some cases we determined $b_2^{\text{cr}}$ only starting from certain minimal $d$ value.

Propagation of perturbed multipole vector solitons is illustrated in Fig. 3. We solved Eqs. (1) by split-step Fourier method for input conditions $q_{1,2}|_{\xi=0} = w_{1,2}(1 + \rho_{1,2})$, where $\rho_{1,2}(\eta, \zeta)$ stand for white noise with the Gaussian distribution and variance $\sigma_{\text{noise}}^2$. Stable multipole vector solitons retain their structure over indefinitely long distances even in the presence of considerable input noise (see Figs. 3(a) and 3(b)). Similar scenarios were encountered for higher-order vector solitons. Interestingly, the $w_2$ component of weakly unstable multipole solitons in the strongly nonlocal media may undergo noise-induced kaleidoscopic transformations, like those shown in Fig. 3(c), periodically almost restoring its input structure (thus, in Fig. 3(c) first restoration of input intensity distribution occurs at $\xi \approx 540$).

In conclusion, we have analyzed the existence and stability of multipole vector solitons in media with Helmholtz-type nonlocal nonlinear response. We revealed that in such media vectorial coupling with the nodeless beam is a necessary condition for stabilization of multipole solitons, which in the scalar case are highly unstable upon propagation. Our findings suggest that the physical nature, hence the spatial shape of nonlinear response function, is a crucial factor for stability of higher-order solitons.

*Also with Universidad de las Americas, Puebla, Mexico. **Also with National Institute of Physics and Nuclear Engineering, Bucharest, Romania.

# References with titles

# Figure captions

Figure 1. Total refractive index profile (left column) and field modulus distribution in second component (right column) for (a) dipole soliton at $b_1 = 3$, $b_2 = 1.8$, and $d = 3$, (b) quadrupole soliton at $b_1 = 3$, $b_2 = 0.9$, and $d = 4$, (c) hexapole soliton at $b_1 = 3$, $b_2 = 0.4$, and $d = 5$.

Figure 2. (a) Energy flow of dipole soliton vs propagation constant $b_2$ at $d = 1$ (1) and 0.2 (2). (b) Domain of existence of dipole solitons on the plane $(d, b_2)$. (c) Critical propagation constant vs nonlocality degree for dipole solitons. (d) Energy sharing between components of dipole soliton vs propagation constant $b_2$. (e) Domain of existence of quadrupole solitons on the plane $(d, b_2)$. (f) Critical propagation constant vs nonlocality degree for quadrupole solitons. In all cases $b_1 = 3$.

Figure 3. Propagation dynamics of vector solitons in the presence of white input noise with variance $\sigma_{\text{noise}}^2 = 0.01$. (a) Dipole soliton at $b_1 = 3$, $b_2 = 1.8$, and $d = 3$. (b) Quadrupole soliton at $b_1 = 3$, $b_2 = 0.9$, and $d = 4$. (c) Hexapole soliton at $b_1 = 3$, $b_2 = 0.4$, and $d = 5$. Only intensity distributions in second components are shown at different distances.

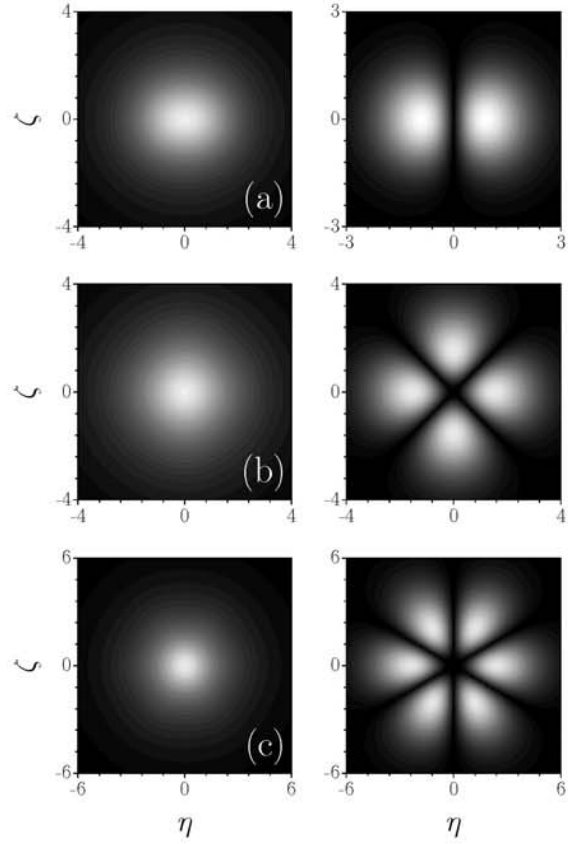

Figure 1. Total refractive index profile (left column) and field modulus distribution in second component (right column) for (a) dipole soliton at $b_1 = 3$, $b_2 = 1.8$, and $d = 3$, (b) quadrupole soliton at $b_1 = 3$, $b_2 = 0.9$, and $d = 4$, (c) hexapole soliton at $b_1 = 3$, $b_2 = 0.4$, and $d = 5$.

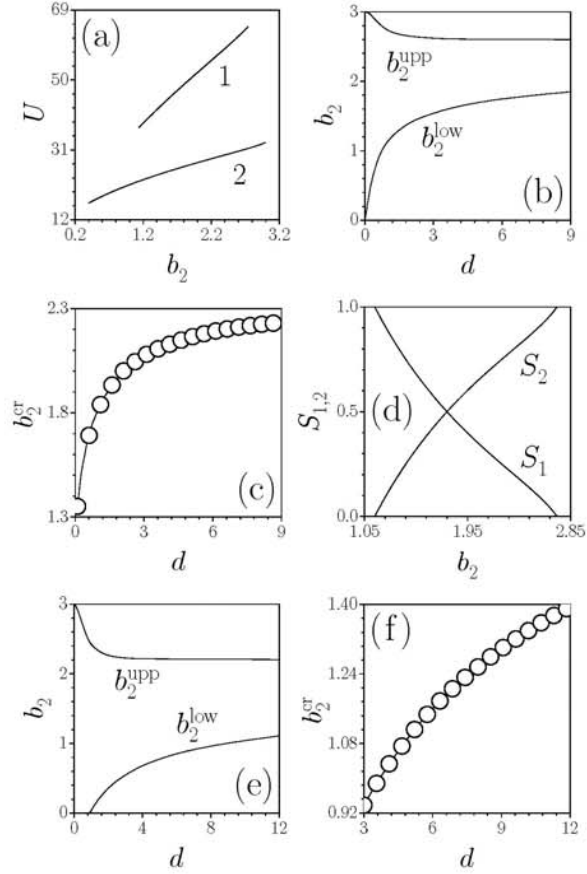

Figure 2. (a) Energy flow of dipole soliton vs propagation constant $b_2$ at $d = 1$ (1) and 0.2 (2). (b) Domain of existence of dipole solitons on the plane $(d, b_2)$. (c) Critical propagation constant vs nonlocality degree for dipole solitons. (d) Energy sharing between components of dipole soliton vs propagation constant $b_2$. (e) Domain of existence of quadrupole solitons on the plane $(d, b_2)$. (f) Critical propagation constant vs nonlocality degree for quadrupole solitons. In all cases $b_1 = 3$.

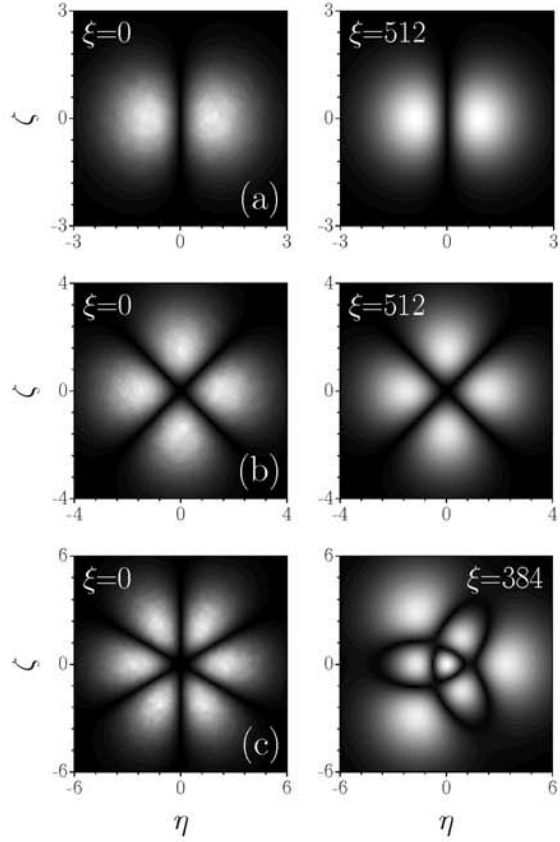

Figure 3. Propagation dynamics of vector solitons in the presence of white input noise with variance $\sigma_{\text{noise}}^2 = 0.01$. (a) Dipole soliton at $b_1 = 3$, $b_2 = 1.8$, and $d = 3$. (b) Quadrupole soliton at $b_1 = 3$, $b_2 = 0.9$, and $d = 4$. (c) Hexapole soliton at $b_1 = 3$, $b_2 = 0.4$, and $d = 5$. Only intensity distributions in second components are shown at different distances.